\documentclass[twocolumn,showpacs,preprintnumbers,amsmath,amssymb]{revtex4}
\usepackage{graphicx}% Include figure files
\usepackage{dcolumn}% Align table columns on decimal point
\usepackage{bm}% bold math
\begin{document}
%\draft

\title{Phase Diagram of Half Doped  Manganites.}

\author{Luis Brey }

\affiliation{\centerline {Instituto de Ciencia de Materiales de
Madrid (CSIC),~Cantoblanco,~28049~Madrid,~Spain.}} \
%\maketitle
\begin{abstract}

An analysis  of the properties of half-doped manganites is presented. We build  up the phase diagram of the
system combining a realistic calculation of the electronic properties and a mean field treatment of the
temperature effects. The electronic structure of the manganites are described with a double exchange model with
cooperative Jahn-Teller phonons and antiferromagnetic coupling between the $Mn$ core spins. At zero temperature
a variety of electronic phases as ferromagnetic (FM) charge ordered (CO) orbital ordered (OO), CE-CO-OO and  FM
metallic, are obtained. By raising the  temperature  the  CE-CO-OO phase becomes paramagnetic (PM), but
depending on the electron-phonon coupling and the exchange coupling the transition can be direct or trough
intermediate states: a FM disorder metallic, a PM-CO-OO  or a FM-CO-OO. We  also discus the nature of the high
temperature PM phase in the regime of finite electron phonon coupling. In this regime half of the oxygen
octahedra surrounding the $Mn$ ions are distorted. In the weak coupling regime the octahedra are slightly
deformed and only trap a small amount of electronic charge, rendering the system metallic consequentially.
However in the strong coupling regime the octahedra are strongly distorted, the charge is fully localized in
polarons and the system is insulator.

\end{abstract}


\maketitle

\section{Introduction}
Oxides of composition (R$_{1-x}$A$ _x$)MnO$_3$ where R denotes rare earth ions and A is a divalent alkaline ion,
are called generically manganites. In these compounds $x$ coincides with the concentration of holes moving in
the $e_g$ orbital band of the Mn ions that ideally form a cubic structure. The electronic and magnetic
properties of manganites are determined by the competition between at least four independent energy scales; the
anti-ferromagnetic interaction between  the  Mn spins, the electron phonon coupling, the electronic repulsion
and the kinetic energy of the carriers. In manganites the energy magnitude of these effects is the same and very
different states can have very similar energies. Consequently slightly varying  parameters as carrier
concentration, strain, disorder, temperature etc, different phases  can be experimentally observed.

In this work we calculate the phase diagram of manganites at half
doping, $x$=1/2. We study the different phases of the system as
function of temperature ($T$), electron-phonon coupling
($\lambda$) and antiferromagnetic superexchange coupling between
the classical $Mn$ $t_{2g}$ core spins ($J_{AF}$). In particular
we are interested in analyzing  the different phase transitions
that undergoes the paramagnetic phase when the temperature is
lowered\cite{Schiffer_1995,Mori_1999}.

Previous temperature phase diagrams have been obtained for the one orbital double exchange
model\cite{Calderon1,Alonso,JAV}. Although this model clarify some physics of the manganites at low hole doping,
the one orbital model is insufficient for describing the complexity of half doped manganites. On the other hand
Monte Carlo simulation of a two orbital double exchange model are very expensive computationally and only very
small systems can be studied\cite{Aliaga}. In this paper the phase diagram is obtained combining the electronic
properties obtained from a  realistic microscopic two-dimensional Hamiltonian and a mean field treatment of the
temperature fluctuations. With this method is possible to analyze system with up to 26$\times$26 Mn ions, and
the results are rather free of finite size effects.  It has been shown extensibely\cite{Dagottobook} that, due
to the directionality of the electronic active $e_g$ $Mn$ orbitals,  the phases appearing in two-dimensional
calculations are qualitatively similar to those appearing in three-dimensional models, and the zero temperature
phase diagrams obtained in two and three dimensions are topologically equivalents.

\begin{figure}
  \includegraphics[clip,width=9cm]{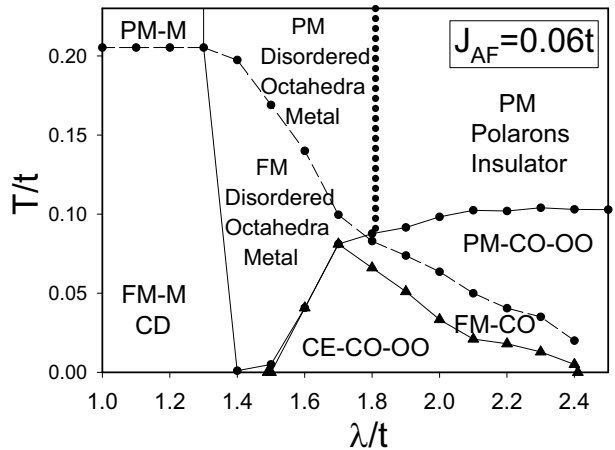}
  \caption{Phase diagram, $T$-$\lambda$, for the two dimensional DE  two orbital model with
  cooperative Jahn-Teller phonons. The AF coupling is
  $J_{AF}$=0.06$t$. The  abbreviations explaining the phases are defined
  in the text. First order transitions are indicated by continuous
  lines and second order transitions by dashed lines. The big dote
  line indicates a metal-insulator transition.
  The PM-M is also CD, and the FM-CO is also OO. Note that
  for 1.4$t < \! \lambda < \!$1.5$t$
  and low $T$ there is a small region, almost imperceptible in the figure,  where the
  system is FM-CO-OO.
  }
\label{fig1}
\end{figure}

\begin{figure}
  \includegraphics[clip,width=9cm]{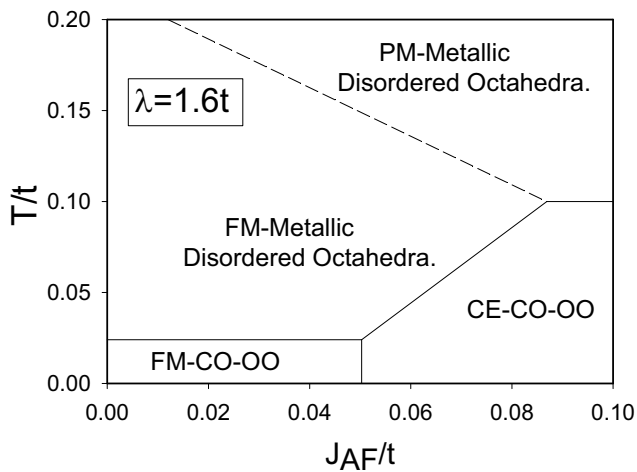}
  \caption{Phase diagram, $T$-$J_{AF}$, for the two dimensional DE  two orbital model with
  cooperative Jahn-Teller phonons. The Jahn-Teller  coupling is
  $\lambda$=1.6$t$. The abbreviations  explaining the phases are defined
  in the text. First order transition are indicated by continuous
  lines and second order transitions by dashed lines.
  }
\label{fig2}
\end{figure}

\begin{figure}
  \includegraphics[clip,width=9cm]{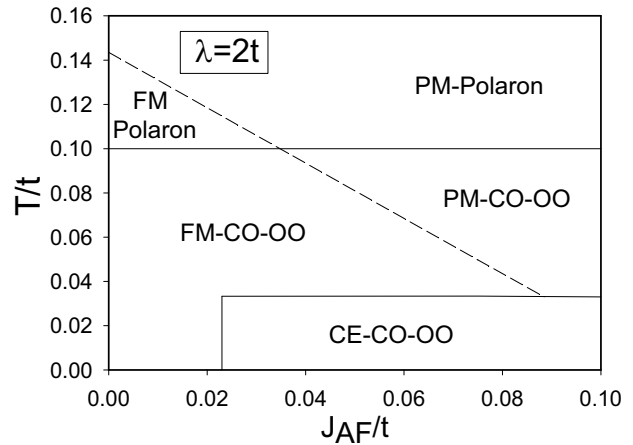}
  \caption{Same than Fig. \ref{fig2} but with $\lambda$=2$t$.}
\label{fig3}
\end{figure}

In Figures \ref{fig1},\ref{fig2} and \ref{fig3} we show the main
results of this work. Fig.\ref{fig1} shows the phase diagram
$T-\lambda$ of the half-doped two-dimensional, two orbital double
exchange model with cooperative Jahn-Teller phonons for
$J_{AF}$=0.06$t$ In Fig.\ref{fig2} and \ref{fig3} we plot the
$T-J_{AF}$ phase diagrams for $\lambda$=1.6$t$ and $\lambda$=2$t$
respectively. Here $t$ is the hopping between first neighbors
$d_{x^2-y^2}$ orbitals in the $x$-$y$ plane, and in the rest of
the paper it is taken as the unit of energy.

The magnetic phases appearing in the diagram are: ferromagnetic
(FM), antiferromagnetic (AF), paramagnetic (PM) and CE
(ferromagnetic zigzag chains that are coupled
antiferromagnetically\cite{Wollan}).
%From the electronic point of
%view,
In some phases the electric charge is chessboard-like ordered (CO phases). The charge modulation is due to the
cooperative Jahn-Teller coupling.  The electronic charge trapped in a $Mn$ ion depends on the distortion of the
oxygen octahedron surrounding the ion, that at the same time depends on $\lambda$. The octahedra distortions of
first neighbors $Mn$ ions are coupled and due to the elastic energy a distorted octahedron is encircled by
undistorted octahedra. The order in the octahedra distortions produces the electron charge modulation. For large
values of the coupling,  $\lambda \ge 2.3t$, the charge modulation is maximum and the CO phase is an ordered
distribution of $Mn^{3+}$ and $Mn^{4+}$. However  for weaker values of $\lambda$ the modulation is much smaller.
The cooperative nature of the Jahn-Teller coupling implies that the CO is concomitant to the existence of
orbital ordering (OO). Charge disordered (CD) names metallic phases (M) where the electric charge moves freely
between the $Mn$ ions and is equally distributed between them. For strong Jahn-Teller coupling, $\lambda \geq
1.9t$, by raising $T$ the CO phase melts into an insulator polaron-like state where the carriers  are
practically fully localized in $Mn$ ions surrounded by a strong distorted oxygen octahedron. For weak values of
$\lambda$ the CO phase melts into a state formed  of weak deformed oxygen octahedra, that trap only a small
amount of the electronic charge. This phase is metallic and can be considered as a \emph{crossover} between the
normal Fermi liquid phase and the polaron-like state\cite{Ciuchi_1999,Blawid_2000}. In the phase diagrams,
Figures \ref{fig1},\ref{fig2} and \ref{fig3}, the solid and the dashed lines represent first and second order
phase transitions respectively.  In Fig \ref{fig1}  the large dot line separating the disordered octahedra
metallic phase from the polaron-like state, is an estimation  of the value of $\lambda$ where the system
undergoes a metal insulator transition. Some of the high temperature phases appearing in  Figures
\ref{fig1},\ref{fig2} and \ref{fig3} have been not reported in Monte Carlo simulations. We believe that due to
size effects the disordered octahedra phases could  be difficult to identified  in small systems.

Due to computational limitations we have only considered spatially
uniform phases in Figures \ref{fig1},\ref{fig2} and \ref{fig3} .
Therefore we can not rule out the existence of more sophisticated
spatially commensuarte/inconmensurate modulated
phases\cite{Chen_1996,Kajimoto_2001,Mathur,Milward,Brey_2004,Loudon2}.
In the calculation we have not considered coupling between
different order parameters. This could affect or phase diagram
near crossing between different lines of phase transitions. Also
near the crossing points the thermal fluctuations that are
neglected in the mean field approximation could be important.

Some characteristics  of the phase diagrams explain several
experimental results on manganites and deserves to be discussed
explicitly:
\par \noindent
i)There is a region on the parameter space, small $J_{AF}$ large
$\lambda$, where the zero temperature ground state is FM and
presents charge and orbital ordering, FM-CO-OO. This phase results
from the cooperative JT phonon and does not requires a strong
$J_{AF}$ coupling to exists\cite{Hotta1,Aliaga}. A phase with the
same symmetry has been observed recently by Loudon {\it et
al.}\cite{Loudon1}.
\par\noindent
ii)In agreement with several experimental results\cite{Goodenough,
Radaelli_1997,Tomioka_2002,Rivadulla_2002,Aladine_2002,Greiner_2004}
and theoretical
works\cite{Solovyev,Brink,Dagottobook,Ferrari,Calderon2} there is
a wide region of the phase diagram (intermediate values of
$\lambda$ and $J_{AF}$), where the low temperature ground state is
the insulator CE-CO-OO phase.
\par\noindent
iii)At high temperature the system is always paramagnetic, but
depending on the value of the electron phonon coupling the system
behaves as a metal(low $\lambda$) a disorder metal (moderates
values of $\lambda$) or  an insulator (strong coupling). These
regimes have been observed experimentally. In the case of
La$_{0.5}$Sr$_{0.5}$MnO$_3$\cite{Urushibara_1995} the high
temperature paramagnetic phase is metallic, whereas in perovskites
with strong electron-phonon coupling, as
La$_{0.5}$Ca$_{0.5}$MnO$_3$\cite{Schiffer_1995} the paramagnetic
phase is an insulator.
\par
\noindent iv)By raising the temperature  the CE-CO-OO insulating
phase becomes a PM phase, but depending on the values of $\lambda$
and $J_{AF}$, in the heating process the transition is trough a FM
disorder metallic phase or through a PM-CO-OO phase. These phase
transition paths  describe appropriately the experimental
phenomenology. Upon cooling, La$_{0.5}$Ca$_{0.5}$MnO$_3$ first
becomes  FM-CO and then CE\cite{Schiffer_1995,Rivadulla_2002a},
however in the case of Pr$_{0.5}$Ca$_{0.5}$MnO$_3$, the
intermediate phase is a PM-CO phase\cite{Mori_1999,Kajimoto_2001}

This article is organized as follows. In Sec.II we describe the
model and the approximations used to find the energies and
properties of the different phases. In Sec.III, we describe the
phases studied and outline the method for obtaining the critical
temperatures and the transitions between the phases. In Sec.IV we
summarize the results.

\section{Model.}
We assume that the manganites crystallize in an ideal perovskites
structure where the Mn ions form a cubic lattice. The crystal
field splits the Mn $d$ levels into an occupied strongly localised
$t_{2g}$ triplet and a doublet of $e_g$ symmetry. The Coulomb
interaction between electrons prevents double occupancy and aligns
the spins of the $d$ orbitals.
%At $x$=1 the $e_g$ orbitals are empty and the
%superexchange coupling between the Mn  spins produces an AF GS.
At $x \neq $1 there is a finite density of electrons in the system
that hop  between the empty $e_g$ $Mn$ states. The Hund's coupling
between the spins of the carriers and each core spin is much
larger than any other energy in the system, and each electron spin
is forced to align locally with the core spin texture. Then the
carriers can be treated as spinless particles and the  hopping
amplitude between two Mn ions is modulated by the spin reduction
factor,
\begin{equation}
f_{12}= \cos\frac{\vartheta _1}{2}\cos\frac{\vartheta _2}{2} + e
^{i ( \phi _1 - \phi _2)} \sin\frac{\vartheta
_1}{2}\sin\frac{\vartheta _2}{2} \label{SRF}
\end{equation} where
$\{\vartheta _i, \phi _i \}$ are the Euler angles of the, assumed
classical, Mn core spins $\{\textbf{S} _i \}$ . This is the so
called  DE model\cite{Zener,Anderson,DeGennes}.

For obtaining the states  of the system we study a two dimensional
DE model coupled to Jahn-Teller (JT) phonons. We also include the
AF coupling between the Mn core spins $J_{AF}$.
\begin{eqnarray}\label{Hamiltonian}
 H &  = &  -\sum_{i,j,a,a'} f _{i,j} t _{a,a'} ^{u} C ^+ _{i,a} C
_{j,a'}
\nonumber \\
 &+& J_{AF} \sum _{<i,j>}
\textbf{S} _i \textbf{S} _j
 + \frac{1}{2} \sum _i \left ( \beta Q _{1i} ^2+ Q
^2_{2i} + Q _{3i} ^2 \right )
\nonumber \\
 & + & \lambda \sum _{i} \left ( Q _{1i} \rho _i + Q _{2i} \tau _{xi}
+ Q _{3i} \tau _{zi} \right ) \, \, \, ,
\end{eqnarray}
here $C^+ _{i,a}$ creates an electron in  the Mn ions located at site $i$ in the $e_g$ orbital $a$ ($a$=1,2 with
1=$|x^2-y^2>$ and 2=$|3z^2-r^2>$). The hopping amplitude is finite for next neighbors  Mn and depends both on
the type of orbital involved and on the direction $u$ between sites $i$ and $j$ ($t_{1,1}^{x(y)}=\pm \sqrt{3}
t_{1,2}^{x(y)} =\pm \sqrt{3} t_{2,1}^{x(y)}=3t_{2,2}^{x(y)}=t)$\cite{Dagottobook}. $t$ is taken as the energy
unit.  The forth term couples the $e_g$ electrons with the three active MnO$_6$ octahedra distortions: the
breathing mode $Q_{1i}$, and the JT modes $Q_{2i}$ and $Q_{3i}$ that have symmetry $x^2$-$y^2$ and $3z^2$-$r^2$
respectively. $Q_{1i}$ couples with the charge at site $i$,  $\rho _i = \sum _a C^+ _{i,a} C _{i,a}$ whereas
$Q_{2i}$ and $Q_{3i}$ couple with the $x$ and $z$ orbital pseudospin, $\tau _{xi}= C^+ _{i1}C_{i2} + C^+ _{i2}
C_{i1}$ and $\tau _{zi}= C^+ _{i1}C_{i1} - C^+ _{i2} C_{i2}$, respectively.  The third term is the elastic
energy of the octahedra distortions, being $\beta \geq$2 the spring constant ratio for breathing and
JT-modes\cite{Aliaga}. In the perovskite structures the oxygens are shared by neighbouring MnO$_6$ octahedra and
the $Q$'s distortions are not independent, cooperative effects being  very important\cite{JAV}. In order to
consider these collective effects we consider the position of the oxygen atoms as the independent variables of
the JT distortions.

For a given value of the parameters $ \lambda$ and  $J_{AF}$, and a texture of core spins $\{ \textbf{S}_i \}$,
we solve self-consistently the mean field version of Hamiltonian (\ref{Hamiltonian}) and obtain the energy, the
local charges $\{\rho_i\}$, the orbital pseudospin order $\{\tau _{xi}, \tau_{zi} \}$ and the oxygen octahedra
distortions $Q_{\alpha,i}$. These quantities are better described by their Fourier transforms, that are
represented by the same symbol with a hat: $\hat{\rho} (\bf{G})$, $\hat{Q _1} (\bf{G})$, ... We have checked
that with our model and the appropriated parameters we recover some well known states as the FM and the
CE-CO\cite{Dagottobook,Aliaga,Brink}. Therefore we are in a position for studying  other phases that can appear
at half doping.

\section{Electric and Magnetic Phases and Critical Temperatures.}

\begin{figure}
  \includegraphics[clip,width=9cm]{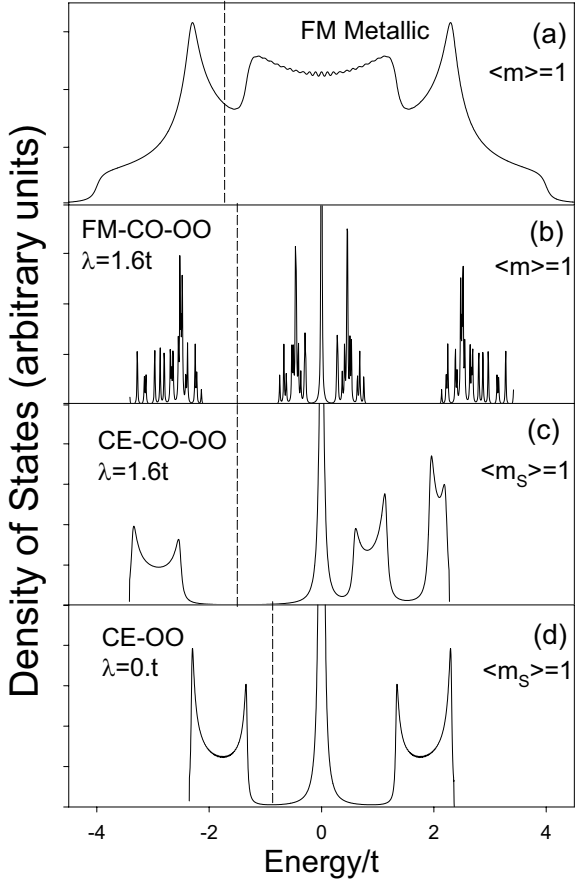}
  \caption{Density of states for different
  ordered phases (see text). The dashed lines indicate the position
  of the Fermi energy. The density of states of the FM-CO-OO
  phase is discrete as it is obtained by solving numerically a
  system with  24x24 $Mn$ ions.}
\label{fig4}
\end{figure}

\subsection{Ferromagnetic Metallic (FM-M) phase.}
In this phase the $Mn$ spins point, on average, in a particular direction, and there is a finite relative
magnetization $< \! m \!
>$. In the spirit of the virtual crystal approximation, we
consider a unique value  for the spin reduction factor $f_{ij}$
that corresponds to its expectation value,
\begin{equation}
f_{ij}\simeq <  \! \! \sqrt{\frac{1 \! + \! \cos{\theta_{ij}}}{2}}
\!
> \! \simeq  \! \sqrt{\frac{1 \! + \! < \!
\cos{\theta_{ij}} \! >}{2}} \! = \! \sqrt{\frac{1 \! +  \! \! < \!
\! m \! \!
>^2}{2}}, \label{srf_ap}
\end{equation}
where $< \, \, \, >$ means thermal average, $\theta _{ij}$
represents the angle formed by the spins located at sites $i$ and
$j$. In Eq.(\ref{srf_ap}) we have approached the expectation value
of the square root by the square root of the average
value\cite{Arovas-1999}. This phase occurs for small values of
$J_{AF}$ and $\lambda$ for  which the octahedra are not distorted
and the system is metallic. In Fig.\ref{fig4}(a) we plot the
density of states (DOS) of the FM metallic phase for $<m>$=1. For
other values of the magnetization the bandwidth is $8t
\sqrt{\frac{1+<m>^2}{2}}$. The internal energy is the sum of an
electronic energy and the AF superexchange energy,
\begin{equation}
E^{FM-M}= \varepsilon _{FM} \sqrt{\frac{1+<m>^2}{2}}+ 2 J_{AF} <m>
^2 \, \, \, . \label{energy_FM}
\end{equation}
Here $\varepsilon _{FM}$ is the electronic energy per $Mn$ ion in
the full polarized ($<m>$=1)  ferromagnetic metallic phase. The
paramagnetic phase corresponds to a random orientation of the $Mn$
spins, $<m>$=0. Both the ferromagnetic and paramagnetic metallic
phases do not present charge ordering, however in order to
minimize the kinetic energy ,the $x^2$-$y^2$ orbital is more
occupied than the $3z^2$-$r^2$ orbital\cite{Aliaga} and these
phases show orbital ordering. The occurrence of  orbital order in
a metallic phase only happen in two dimensions and therefore,
here, we do not consider it as an relevant effect.

In order to describe thermal effects we have to compute the free energy. In the manganites, the typical critical
temperatures are lower than 300$K$ and the thermal energy is much smaller than the electronic energies, $t
\sim$0.2$eV$, and neglect thermal effects on the carriers Fermi distribution is a good approximation. Therefore
we only need to consider the entropy of the classical $Mn$ spins. We use the molecular field approximation,
neglecting all correlations between different spins and assuming for each of them a statistical distribution
corresponding to an effective magnetic field $h$\cite{DeGennes}. The free energy of  a system of classical spins
in presence of $h$ is
\begin{equation}
{\cal F} = -  \frac{1}{\beta} \log \left ( 2 \frac{ \sinh(\beta h
)}{\beta h } \right ), \label{Mn_free_energy}
\end{equation}
and the  magnetization of this system is
\begin{equation}
< \! m \! > = - \frac{\partial {\cal{F}} }{\partial h } =
\frac{1}{\tanh ( \beta h)} -\frac{1}{\beta h}.
\label{magnetization}
\end{equation}
This equation gives the relation between the effective magnetic
field  and the magnetization. The entropy of the spin system is
given by
\begin{equation}
-T S = {\cal{F}} - < \! m \! > h = \frac{1}{\beta} \left ( - \log
\left ( 2 \frac{\sinh ( \beta h )}{\beta h } \right )  - m \beta h
\right ), \label{entropy}
\end{equation}
that can be written as a function of $< \! m \! >$ solving
selfconsistently Eq.\ref{Mn_free_energy} and
Eq.\ref{magnetization}. In the limit of small magnetization the
entropy takes the form,
\begin{equation}
S ( <\! m \! >) =\frac{\log 2}{2}-\frac{3}{2} <\! m \!> ^2-
\frac{9}{20} <\! m \!> ^4 + .... \label{expansion}
\end{equation}
Using this expression for the $Mn$ spins entropy the total free
energy of the system near the PM to FM transition takes the form,
\begin{eqnarray}
F( <\! \!  m \! \! > )& = & E^{FM-M}-  \! T S ( <\! \! m  \! \!>) \nonumber \\
& \approx&
 F(0)+ <\! \! m \! \!
> ^2 (\frac{3}{2} \, T  +  \frac{\varepsilon
_{FM}}{2 \sqrt{2}} + 2 J _{AF} )   \nonumber
\\  & &  \, \, \, \, \, \, \, \, \, \, \, \, \, + <\!  \! m  \! \!
> ^4 ( \frac{9}{20} \,  T - \frac{\varepsilon _{FM}}{8 \sqrt{2}} )+
... \label{freeenergy_total}
\end{eqnarray}
and the Curie temperature is
\begin{equation}
T_C= -\frac{\varepsilon _{FM}}{3 \sqrt{2}}  - \frac{4}{3} J_{AF}
\, \, \, .
\end{equation}
The constant multiplying the $<\! m \! > ^4 $ term is positive and
therefore  the transition is a second order phase transition.

\begin{figure}
  \includegraphics[clip,width=9cm]{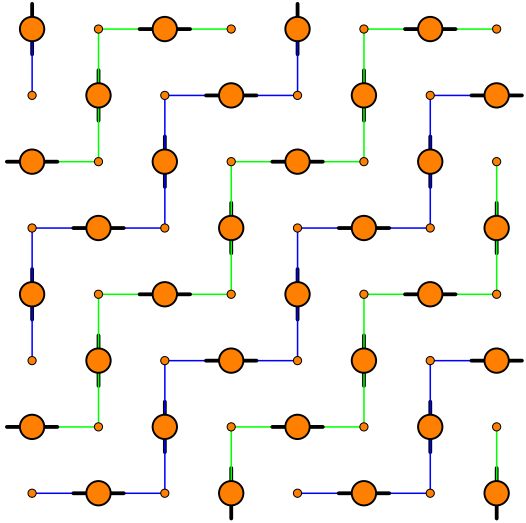}
  \caption{(Color online) Schematic view of CO-OO phases. The
  horizontal  and vertical lines on the lattice sites represent the positive and
  negative oxygen octahedra distortions. The size of the dots is proportional to the charge located on the
  $Mn$ ions. In the case of $\lambda$=1.6$t$ the octahedra
  distortion is $Q_2$=$\pm$1.25$t$ and the ions have charge 0.71 and 0.29.
   In the FM phase all ions have the same spin orientation,
   whereas the CE phase consists of FM  zigzag chains  which are coupled AF.}
  \label{fig5}
\end{figure}

\subsection{Ferromagnetic, Charge Ordered,  Orbital Ordered  phase (FM-CO-OO).}
For small values of $J_{AF}$ and strong enough electron phonon
coupling, $\lambda \gtrsim $1.4$t$, the ground state present
charge and orbital order and a finite relative magnetization,
$<m>\neq 0$. This is a surprising state  as the charge order (CO)
is normally associated with AF order. In manganites the CO is
created by the electron phonon coupling and for realistic values
of $\lambda$, the charge disproportionation between the two
different types of $Mn$ ions is small. Then  the system can reduce
its energy by ordering ferromagnetically the $Mn$ core spins  and
minimizing  the kinetic energy.

The charge ordering is characterized  by a single Fourier
component, ($\pi,\pi$),  of the charge density, $\hat{\rho}
(\pi,\pi)\neq 0$. The orbital ordering is described by a uniform
$z$-orbital pseudospin, $\tau_{z,i}$ and  two finite Fourier
components of the $x$-component of the pseudospin, $\hat{\tau
_x}(\pi/2,\pi/2)=\hat{\tau _x}(-\pi/2,-\pi/2)\neq 0$. Due to the
electron phonon coupling, the amplitude of the charge modulation
is related with the distortion of the oxygen octahedra, $\hat{Q
_2}(\pi/2,\pi/2)=\hat{Q _2}(-\pi/2,-\pi/2)\neq 0$, that  depends
on the value of $\lambda$.  The sites where the electron charge is
larger than the average, $x$=$1/2$, the $Q_2$ mode has a finite
value, whereas the octahedra surrounding $Mn$ ions with less
charge than the average are undistorted. The sign and amplitude of
the distortion are modulated in order to minimize the elastic
energy of the cooperative Jahn-Teller distortion. The CO-OO phase
can be described as a distribution of deformed octahedra. The
position of the distorted octahedra determines  the electron
charge ordering, and the orientation of the octahedra distortions
fix the orbital ordering in the system. CO and OO are a
consequence of the spatial modulation of the octahedra
deformations. In Fig. \ref{fig5} we plot the arrangement of the
octahedra distortion and the electronic charge distribution for a
unit cell with 8x8 $Mn$ ions for $\lambda$=1.6$t$. As the $Mn$
ions have two orbital per site, the $(\pi/2,\pi/2)$ modulation of
the octahedra distortion opens a gap at the Fermi energy for
$x$=1/2 and therefore the system is an insulator. The magnitude of
the gap increases with the value of $\lambda$. In
Fig.\ref{fig4}(b) we plot a typical density of this phase for
$\lambda$=1.6$t$.

A decrease of the relative magnetization $< \! m \! >$ produces a
reduction  in the spin reduction factor $f_{i,j}$ and in the
kinetic energy, making the relative electron-phonon interaction
stronger. The internal energy per $Mn$ ion of this phase can be
written as
\begin{equation}
E^{FM-CO}= \varepsilon _{FM-CO} (\lambda,< \! m \! >) + 2 J_{AF} <
\! m \! > ^2 \, \, \, . \label{energy_FM_CO}
\end{equation}
where the electronic energy depends in a complicated way on
$\lambda$ and $<m>$ and has to be obtained numerically by solving
Eq.(\ref{Hamiltonian}).

The Curie temperature of the FM-CO-OO phase can be calculated
similarly to the FM-M case, and we obtain
\begin{equation}
T_C = - \frac{2}{3} \left . \frac{\partial E^{FM-CO} }{\partial
<\! m \!
> ^2}    \right  | _{< \! m \! > = 0} \, \, \, .
\end{equation}
In the case of the FM-CO-OO  phase the derivative has to be obtained numerically. From higher derivatives of the
internal energy with respect to the magnetization we obtain that the transition is second order.

\subsection{CE, Charge Order, Orbital Order phase (CE-CO-OO).}
Magnetically the CE phase consists of ferromagnetic zigzag chains coupled ferromagnetically. The horizontal and
vertical steps of the chain contain three Mn ions. For finite values of the electron phonon  coupling the
charge, orbitals  and  octahedra distortions are ordered similarly to in the FM-CO-OO phase. In the case of
$\lambda$=0, the CE  phase does not present charge order, however due to the dependence of the electron hopping
in the spatial direction, the system spontaneously  creates an orbital order similar to that occurring in  the
FM-CO-OO phase, $\hat{\tau _z}(0,0) \! \neq \!$0 and  $\hat{\tau _x}(\frac{\pi}{2},\frac{\pi}{2}) \! =
\!\hat{\tau _x}(-\frac{\pi}{2},-\frac{\pi}{2}) \! \neq \! 0$. Independently of the existence of charge
modulation the $(\frac{\pi}{2},\frac{\pi}{2})$ orbital order opens a gap at the Fermi energy an the half doped
system is an insulator. In the $\lambda$=0 case the gap is just due to the orbital order and the system is
classified as a band insulator\cite{Hotta3}. In Fig. \ref{fig4}(c)and(d), we plot the density of states in the
CE phase for $\lambda$=1.6$t$ and $\lambda$=0 respectively. There is a gap at the Fermi energy that  for
$\lambda$=0  is of the order of the hopping parameter, $t$, and that increases with the electron phonon
coupling.

For each value of $\lambda$ there is a minimal value of $J_{AF}$
for the occurrence  the CE-OO-CO phase. At zero temperature and
for $\lambda$=0 the minimal  AF coupling is $J_{AF} \simeq 0.18t$,
whereas for $\lambda$=1.6$t$ and $\lambda$=2$t$ the minimal   AF
coupling  are 0.05$t$ and 0.023$t$ respectively, Figures
\ref{fig2} and \ref{fig3}.

The magnetization  of the CE phase is  described by  the relative
amount of saturation in each zigzag chain $<m_S>$. In the virtual
crystal approximation the fluctuations are neglected and the
hopping is modulated by the spin reduction factor that is
different along the zigzag FM chain, $f ^ {FM}$ than between the
AF coupled chains,$f ^ {AF}$\cite{DeGennes},
\begin{eqnarray}
  f ^ {FM} &=& \sqrt{\frac{1+<\! m_S \! >^2}{2}} \nonumber \\
  f ^ {AF} &=& \sqrt{\frac{1-< \! m_S \! >^2}{2}}\, \, \, .
\label{srf_CE}
\end{eqnarray}
The internal energy of this phase depends on  $\lambda$, and
$<m_S>$ and can be written as,
\begin{equation}
E^{CE-CO}= \varepsilon _{CE-CO} (\lambda,< \! m_S \! >)  \, \, \,
. \label{energy_CE_CO}
\end{equation}
As each $Mn$ spin core is surrounded by two $Mn$ spins coupled FM
and other two coupled AF, the superexchange energy is zero.

In order to compute the Neel temperature of the CE phases, we
introduce an effective field for each spin sublattice. Taking into
account that both, the magnetization and the effective magnetic
field, have different sign in each sublattice, we obtain the same
expression for the entropy than in the FM phase, but just changing
$<\! m \! >$ by $<\! m \! _ S  >$\cite{DeGennes}. With this the
free energy takes the form,
\begin{eqnarray}
F(<\! m \! _ S  >,\lambda) & = & E^{CE-CO} - T S ( <\! \! m_S \!
\!
>)
\nonumber \\
& \simeq  & F(0)+<\!m \! _ S > ^2 \left ( \frac{3}{2} T + a \right
) \nonumber \\ &+ &<\!m \! _ S > ^4 \left ( \frac{9}{20} T + b
\right ) + ...
\end{eqnarray}
with
\begin{equation}
E^{CE-OO} \thickapprox cte+ a <\! m _S \! > ^2 + b <\! m _S \!
> ^4+ ....
\end{equation}
Due to the symmetry of the CE phase the coefficient $a$ is zero,
and the Neel temperature gets the form
\begin{equation}
T_N ^{CE} = - \frac{20}{9} \, b.
\end{equation}
Numerically, we obtain  $b < 0 $ and therefore there is a finite temperature first order phase transition. As it
is shown in Fig. \ref{fig1}-\ref{fig3}, only  for large values of $\lambda$ and $J_{AF}$ the  CE-OO-OO phase
transforms directly into the PM-CO-OO. At smaller  values of $\lambda$  and $J_{AF}$ the transition to the PM
phase is through a FM-CO-OO (Fig \ref{fig1}) or through a FM metallic disordered octahedra phase (Fig.
\ref{fig2}).

\subsection{Antiferromagnetic CO-OO phase}
For large values of $J_{AF}$, the kinetic energy is zero and the
electron charge is fully localized forming a chessboard-like
pattern of $Mn^{3+}$ and $Mn^{4+}$. For finite values of $\lambda$
the octahedra surrounding the Mn$^{3+}$ ions are distorted
cooperatively and the system presents OO and has the same symmetry
than the others CO-OO phases. The minimal value of $J_{AF}$ for
the occurrence of this phase depends on $\lambda$, but in general
is rather large, and this phase is unlikely  to occur in half
doped manganites, and it in this work is not considered.
%
%The internal energy of this phase depends on $J_{AF}$, $\lambda$,
%and the staggered magnetization $<m_{AF}>$ and can be written as,
%\begin{equation}
%E^{AF-CO}= \varepsilon _{AF-CO} (\lambda,<m_{AF}>) - 2 J_{AF}
%<m_{AF}> ^2 \, \, \, . \label{energy_CE_CO}
%\end{equation}

\subsection{Disordered octahedra phases: Polaron phase and metallic \emph{crossover} phase.}
At high temperature the CO phases melt. The nature of the disorder high temperature   phase depends on the
strength of the electron phonon interaction. The melting of a charge density wave is a complicated problem, and
only recently it has been possible to apply an unique treatment for studying the  weak  and the strong coupling
regimes\cite{Ciuchi_1999,Blawid_2000}. Part of the difficulty in studying the melting of a charge density wave
is to define the order parameter that vanishes in the disordered phase. In the manganites this problem can be
avoided  by considering the distribution of the distorted oxygen octahedra instead of the modulation of the
charge density. In the CO-OO phases there are two order parameters, $\xi$, describing the order-disorder
transition in the the spatial distribution of the distorted oxygen octahedra and $\eta$, parameterizing their
orientational order. The octahedra distorted phase can exists on a FM or PM $Mn$ spins background. However, we
have not found any region in the parameter space where the stable phase contains a disorder distribution of
octahedra  on a magnetic CE background.

\begin{figure}
  \includegraphics[clip,width=9cm]{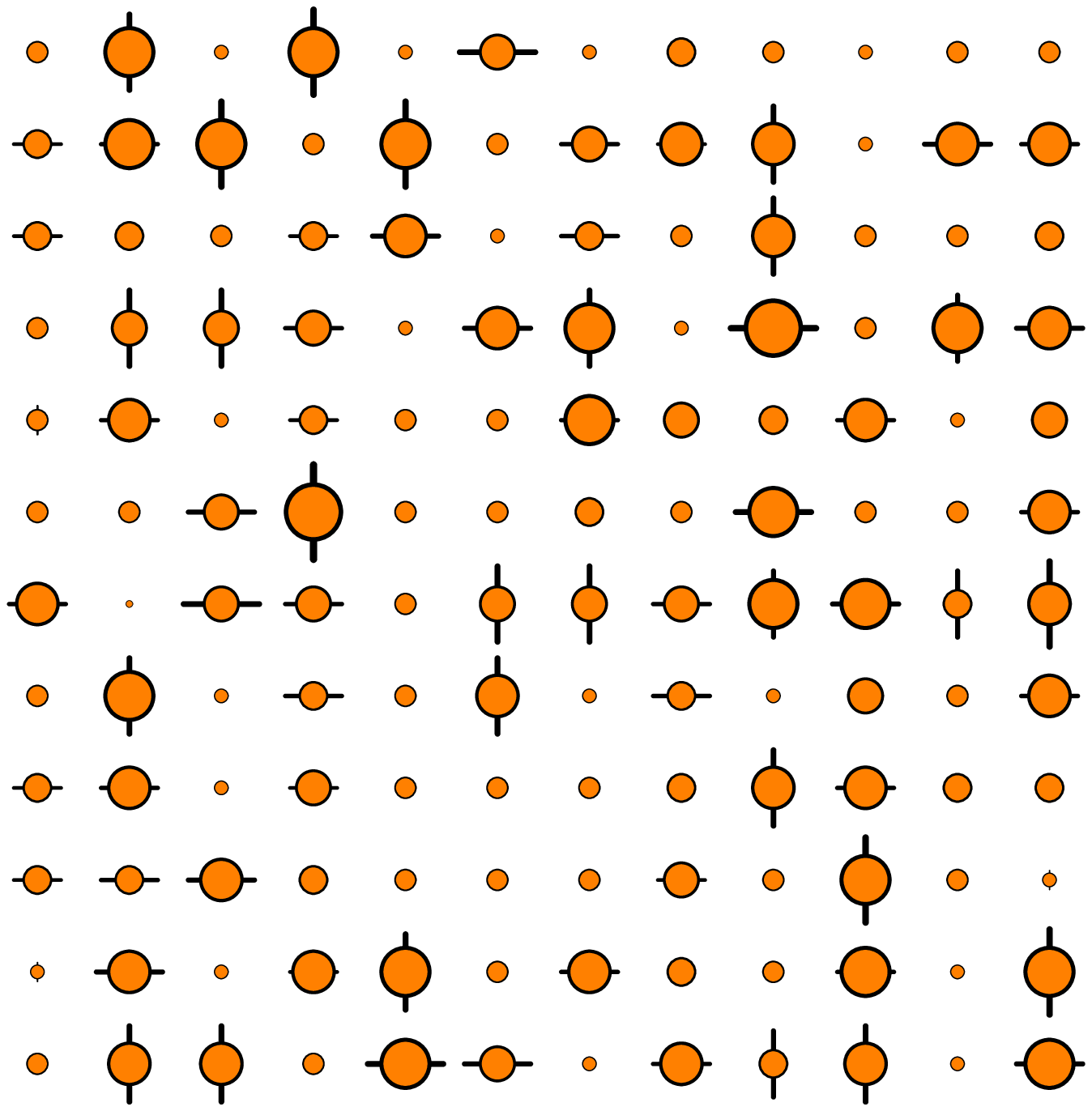}
  \caption{(Color online) Schematic view of the disordered octahedra phase for moderate values
  of $\lambda$. The
  horizontal  and vertical lines on the lattice sites represent the positive and
  negative oxygen octahedra distortions. The size of the dots is proportional to the charge located on the
  $Mn$ ions. For $\lambda$=1.6$t$ the octahedra distortion is
  $Q_2$=$\pm$0.95$t$, and the charge located on the $Mn$ ions is in the range
  0.25-0.75.}
\label{fig6}
\end{figure}

The free energy of the CO-OO phases has the form,
\begin{equation}
F(\xi, \eta) = F (0,0) - T \left (  S _M ( \xi ) + \frac{1}{2} S_M
( \eta) \right ) + E ^{CO-OO} (\xi, \eta) \, \, ,
\end{equation}
being $S _M$ the entropy of mixing\cite{chaikin_book},
\begin{equation}
S _M ( x) = \ln 2 - \frac{1}{2}(1+x) \ln(1+x)- \frac {1}{2} (1-x)
\ln(1-x) \, \, ,
\end{equation}
and $E^{CO-OO} ( \xi, \eta)$ the electronic internal energy measured with respect to the full disordered case.
Near $\xi$=0 and $\eta$=0, the internal energy can be expanded as
\begin{equation}
E^{CO-OO} ( \xi,\eta) = \sum _{n,m} a_{n,m} \xi ^n \eta ^m \, \,
\,
\end{equation}
we have evaluated numerically the lower coefficients in this
expansion and we have found a strong coupling between the two
order parameters that indicates the first order nature of the
order disorder  transition. Assuming a transition between the
fully disordered phase and the fully ordered phase the melting
critical temperature of the CO-OO phases is
\begin{equation}
T_{M} = - \frac{2}{3 \ln 2} \left ( E^{CO-OO} (1,1) - E ^{CO-OO}
(0,0) \right )
\end{equation}

In order to compute the energy of the disordered octahedra phase we locate randomly the distorted octahedra
($\xi$=0). The sign of the distortion is also random ($\eta$=0). We assume that all the octahedra have the same
elongation amplitude, $Q_2$, and we find its value by minimizing the energy. For a given value of the electron
phonon coupling the value of the deformation $Q_2$ is smaller in the disorder phase than in the CO-OO phases. We
have checked in several cases that the energy is minimized when the number of distorted octahedra and the number
of electrons is the same.

\begin{figure}
  \includegraphics[clip,width=9cm]{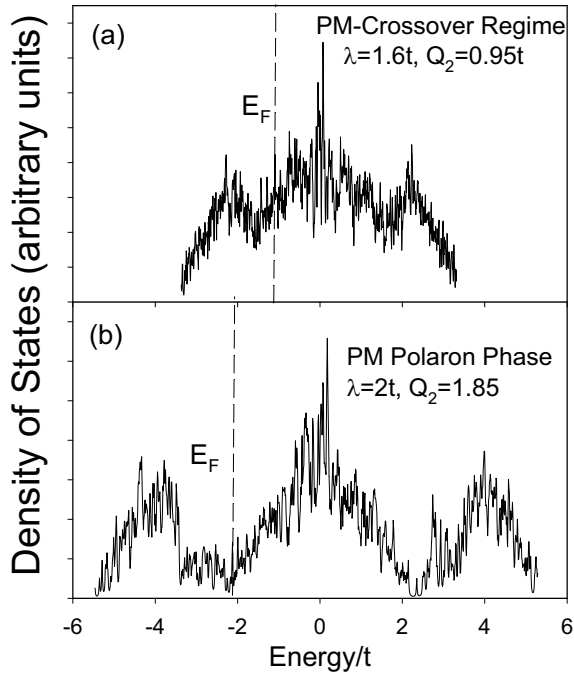}
  \caption{ Density of states of  the disordered octahedra phase
  in the case of (a)  $\lambda$=1.6$t$  and
  (b) $\lambda$=2$t$.
  The dased lines indicate the position
  of the Fermi energy. The density of states is obtained by solving numerically a
  system with  24x24 $Mn$ ions per unit cell.}
\label{fig7}
\end{figure}

\begin{figure}
  \includegraphics[clip,width=9cm]{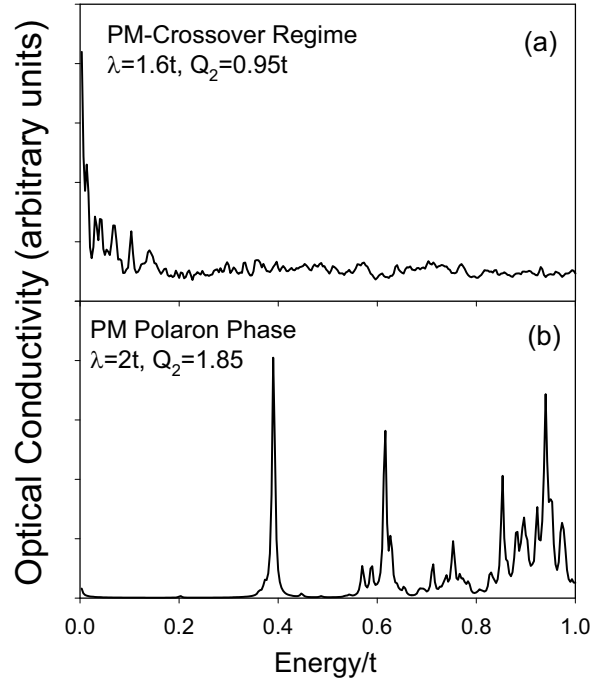}
  \caption{ Optical conductivity
  of  the disordered octahedra phase
  in the case of (a)  $\lambda$=1.6$t$  and
  (b) $\lambda$=2$t$. The optical conductivity is obtained by solving numerically a
  system with  24x24 $Mn$ ions per unit cell.}
\label{fig8}
\end{figure}

By changing the value of $\lambda$, the octahedra disordered phase
has different behaviors. There is a critical value of $\lambda$
where the distortion $Q_2$ become finite. As can be observed in
Fig.\ref{fig1}, this value, $\lambda \sim$1.3$t$, is slightly
smaller than the critical $\lambda$ for the zero temperature FM to
CO-OO transition ($\lambda \sim 1.4 t $ for $J_{AF}$=0.06$t$).
This decrease  occurs because the spin reduction factor is
smaller, and the relative electron phonon coupling larger, in the
PM phase than in the FM phase. For values of $\lambda$ larger than
the critical value, but not too strong, the mode $Q_2$ is small
and the distorted octahedra only trap a small amount of electronic
charge, see Fig.(\ref{fig6}). Interestingly the disorder in the
position and orientation of the Mn octahedra creates states in the
energy gap region of the insulating ordered phase that favor
transport of charge making the system to behave as a
metal\cite{Motome_2003,Sen_2004}. In Fig.\ref{fig7}(a) we plot the
density of states  for a realization of the disordered octahedra
phase for $\lambda$=1.6$t$ in a system containing 24$\times$24
$Mn$ ions. The amplitude of the distortion that minimizes the
energy is $Q_2$=0.95. There is not gap in the DOS, and there is a
finite number of disorder induced states at the Fermi energy. It
is interesting to compare this DOS with that of the ordered phases
calculated with the same $\lambda$, Fig. \ref{fig4}(c)-(d). The
energy gap of the ordered phase disappears when disorder in the
positions and orientation of the octahedra is introduced. The
states at the Fermi energy transport electric charge and the
metallic behavior of this phase is reflected in the low frequency
spectral weight of the optical conductivity $\sigma (\omega)$,
Fig.\ref{fig8}(a), that shows a clear Drude-like character.

\begin{figure}
  \includegraphics[clip,width=9cm]{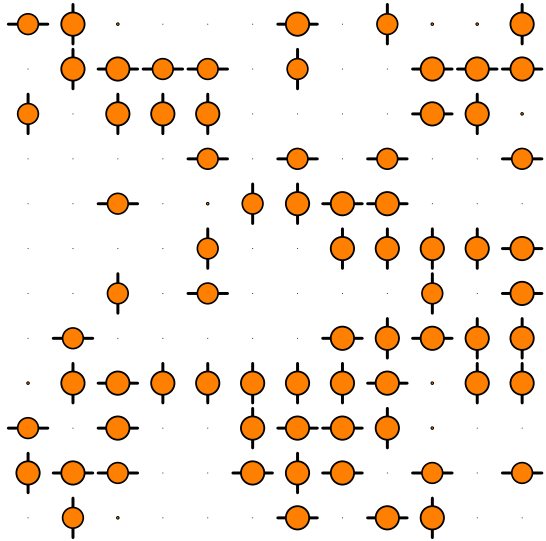}
  \caption{(Color online) Schematic view of the disordered octahedra phase for large values
  of $\lambda$. The
  horizontal  and vertical lines on the lattice sites represent the positive and
  negative oxygen octahedra distortions. The size of the dots is proportional to the charge located on the
  $Mn$ ions. For $\lambda$=2$t$ the octahedra distortion is
  $Q_2$=$\pm$1.85 $t$, and the electrons are practically fully localized in the $Mn$ ions surrounded
  by distorted octahedra.}
\label{fig9}
\end{figure}

By increasing $\lambda$ the octahedra distortion increases, the
electronic charge becomes strongly localized, and the system is a
disordered mixing of Mn$^{3+}$ and Mn$^{4+}$. We name this state
polaronic phase. In Fig.\ref{fig9} a typical octahedra
distribution and electron charge in the polaronic regime,
$\lambda$=2$t$, is plotted. Note that the electron charge is
practically localized on $Mn$ ions surrounded by a distorted
oxygen octahedron. In this strong coupling regime the DOS shows a
fully occupied polaron band an a strong suppression of states at
the Fermi energy as can be observed in Fig. \ref{fig7}(b). This is
an indication of the insulator behaviour of this phase, being more
evident in the optical conductivity Fig.\ref{fig8}(b), that shows
a strong reduction of the low energy weight and the development of
a gap structure.

Although is very difficult to calculate numerically the critical
value of  $\lambda$ for the metal insulating transition in the
disordered octahedra phase, by comparing  qualitatively  the DOS
and the optical conductivity we estimate, $\lambda \sim$1.5$t$, as
the critical value of the electron phonon coupling
(Fig.\ref{fig1}).

\section{Summary}
We have calculated the phase diagram of manganites at half doping.
By combining a realistic microscopic electronic model and a mean
field treatment of  thermal effects, we have obtained the free
energy of the different possible states. With these ingredients we
build up phase diagrams as function of  temperature and electron
phonon and as function of temperature and antiferromagnetic
coupling. By changing the value of the electron phonon or the
antiferromagnetic coupling we simulate different manganites with
different controlled  bandwidth.  The calculated phase diagrams
show a big variety of exotic phases some of them, as the CE-CO-OO
and the FM-CO-OO, have been observed experimentally.

We have obtained that the transition from the high temperature paramagnetic phase to the low temperature CE
phase, depends on the parameters $\lambda$ and $J_{AF}$. For large values of $\lambda$ the transition can be
through a FM-CO-OO phase (small $J_{AF}$)  or trough a PM-CO-OO phase (large $J_{AF}$). However for small values
of $\lambda$ the transition can be direct (large $J_{AF}$) or trough a disorder FM phase (small $J_{AF}$).

We have also discussed the nature of the high temperature PM phase
in the regime of finite $\lambda$. We have obtained that in the
weak coupling regime the oxygen octahedra surrounding the $Mn$
ions are slightly distorted and only trap a small amount of
electronic charge, being the system  metallic. However in the
strong coupling regime half of the octahedra are strongly
distorted and the charge is localized in the $Mn$'s ions encircled
by a distorted octahedron, forming a polaron and making the system
insulator.

\section{Acknowledgments}
The author thanks M.J.Calder\'on, P.B.Littlewood, N.D.Mathur and
P.L\'opez-Sancho  for helpful discussions. This author thanks
Cambridge University for hospitality during the realization of
this work. Financial support is acknowledged from Grants No
MAT2002-04429-C03-01 (MCyT, Spain), Fundaci\'on Ram\'on Areces and
Secretaria de Estado de Universidades. The work at Cambridge was
partially supported by the UK EPSRC.

%\bibliography{mia}

\end{document}